\documentstyle{article}

\begin{document}
\begin{center}
{\bf \Large{ SUPERSYMMETRIC TECHNIQUES APPLIED TO THE JACOBI EQUATION}\\
}

Haret C. Rosu\footnote{e-mail: rosu@ifug3.ugto.mx}
and Juan R. Guzm\'an

%e-mail: rosu@ifug3.ugto.mx\\[2mm]
Instituto de F\'{\i}sica - IFUG,
Apdo Postal E-143,
Le\'on, Gto, Mexico
\end{center}
$ $\\[1.5mm]

\bigskip

{\bf Abstract}. The simple supersymmetric approach
recently used by Dutt, Gangopadhyaya, and Sukhatme
%[Am. J. Phys. {\bf 65} 400 (1997)]
for spherical harmonics is generalized to Jacobi equation, including also
the intermediate Gegenbauer case.

\bigskip

{\bf Resumen}. Un procedimiento supersim\'etrico simple recientemente
usado por Dutt, Gangopadhyaya y Sukhatme
%[Am. J. Phys. {\bf 65}, 400 (1997)]
para los harmonicos esf\'ericos, es generalizado a la ecuacion de
Jacobi, incluyendo tambien el caso intermedio de Gegenbauer.

\bigskip

{\bf European J. Phys. 19 (May 1998) 287-292} [physics/9701016]

%\end{abstract}

\bigskip
\bigskip
\newpage

{\bf I. INTRODUCTION}

During 1971-1974, supersymmetry (SUSY), a more general (graded) Lie algebraic
realization of
symmetries, involving combinations of commutation and anti-commutation
relations has been discovered as a sort of necessary unifying
algebraic ingredient of microscopic bosonic and fermionic degrees of freedom
\cite{s71}.
In 1981, Witten \cite{wit}
introduced the simplest SUSY realization within
nonrelativistic quantum mechanics (SUSYQM) as a toy tool for studying SUSY
breaking in quantum field theories.
SUSYQM may be seen as a special property of the factorizations of the
one-dimensional (1D) Hamiltonians. Namely, one can find a pair of
Hamiltonians
(usually called bosonic and fermionic partners) having the potentials
connected by a single Riccati solution, known as the superpotential.
Indeed, if one fixes to zero the ground state energy,
%then one has the
%following Schr\"odinger equation for the ground state wave
one can easily factorize any 1D Hamiltonian as follows
%%%%%%%%
$$
H_{1}\equiv -c^2\frac{d^2}{dx^2}+V_1=A^{\dagger}A~,
\eqno(1)
$$
%%%%%%%%%%%
where
$
A^{\dagger}=-c\frac{d}{dx}+W(x), A=c\frac{d}{dx}+W(x)
$
%%%%%%%%%%%%
and $c=\frac{\hbar}{\sqrt{2m}}$ for nonrelativistic quantum mechanics.
The $W$ function occurring in the factorizing
operators is the solution (superpotential) of the following Riccati equation
%%%%%%%%%%
$$
V_1=W^2-cW^{'}~.
\eqno(2)
$$
%%%%%%%%%%%%
The SUSYQM scheme, as initiated by Witten, means to construct the ``fermionic"
Hamiltonian obtained by reversing the order of the factoring operators, which
reads
%%%%%%%%%%%
$$
H_{2}=AA^{\dagger}\equiv-c^2\frac{d^2}{dx^2}+V_2~,
\eqno(3)
$$
%%%%%%%%%%
where
%%%%%%%
$$
V_2=W^2+cW^{'}~.
\eqno(4)
$$
%%%%%%%%%%%
One can show that the energy eigenvalues, the wave functions, the $S$
matrices, the reflection and transmission coefficients of the
supersymmetric partners $H_1$ and $H_2$ are related in a simple manner
through the application of the factorizing operators
\cite{rev}. In particular, the energy spectra are identical except perhaps
for the bosonic ground state (for the so-called unbroken SUSY case).
In 1983, Gendenshtein \cite{gen} noticed that in many important problems
there is a very useful relationship
between the SUSY partner potentials $V_{1,2}$, known as shape invariance (SI).
He showed that if the superpartners are similar in shape and differ only
in the parameters that appear in them, i.e., in mathematical terms, if one
writes down a SI condition of the form
%%%%%%%%%%%%%%
$$
V_{2}(x;a_1)=V_{1}(x;a_2)+R(a_1)~,
\eqno(5)
$$
%%%%%%%%%%%%%%
where $a_1$ is a set of parameters, and $a_2$ is a function of $a_1$,
such that the remainder $R (a_1)$ is independent of $x$, one can easily
derive the energy eigenvalues and eigenfunctions of {\em any}
potential fulfilling the SI condition. Thus, SI is a valuable parametric
connection within SUSYQM.

The coordinate transformation procedure is well known in the framework of
SUSYQM (see section 5 in \cite{rev}).
Essentially, one starts with a 1D Schr\"odinger equation and
tries to obtain a better known one by performing a coordinate transformation.
This may be considered as a particular case of various transformations
%which eliminates first order derivatives in
of homogeneous linear differential equations of second order
involving either the dependent variable, or the independent one, or both,
which are recognized as basic methods in the
mathematical literature; the reader is directed to the textbooks of
Szeg\"o and Zwillinger \cite{math}.
However, using SUSYQM, one can exploit
further SUSY symmetries, like the aforementioned SI, to get
some properties
of special functions in an illuminating way. This is precisely the case of
a recent paper of Dutt, Gangopadhyaya and Sukhatme (DGS) \cite{dgs}, where
they used the change of the independent
variable, to recast the associated Legendre equation in
the celebrated Schr\"odinger equation with the $-{\rm sech}$ potential,
which enjoys the SI property. Next, they put to work this concept
to derive some known features of spherical harmonics in a simple way.

In the following, after briefly repeating the DGS case of spherical harmonics,
we generalize their scheme to the Jacobi equation, presenting also the
intermediate Gegenbauer case.
\bigskip
\newpage
{\bf II. SPHERICAL HARMONICS}

The equation for the associated Legendre polynomials
%%%%%%%%%%%%%%%%
$$
\frac{d^2y}{d\theta ^2}+{\rm cot}\theta\frac{dy}{d\theta}+\Big[l(l+1)-
\frac{m^2}{\sin ^2\theta}\Big]y=0
\eqno(6)
$$
%%%%%%%%%%%%%%%%%%
can be transformed into a Schr\"odinger eigenvalue equation by a
mapping function $\theta=f(z)$
that can be found from the condition of putting to nought the coefficient
of the first derivative. The result is $\theta \equiv f=2{\rm arctan}(e^{z})$.
This mapping is equivalent to the replacement $\sin \theta={\rm sech}z$ and
$\cos \theta=-{\rm tanh}z$. The range of the variable $z$ is the full real
line $-\infty <z<\infty$. We notice that
$\theta (z)= \frac{\pi}{2}+{\rm gd} (z)$, where ${\rm gd}(z)$ is the so-called
{\em Gudermannian} or hyperbolic amplitude function \cite{gr}.
The associated Legendre equation is transformed
in one of the best known exactly solvable, SI Schr\"odinger
equation
%%%%%%%%%%%%%%%%%%%%
$$
-\frac{d^2v}{dz^2}-\Big[l(l+1){\rm sech} ^2z\Big]v=-m^2v
\eqno(7)
$$
%%%%%%%%%%%%%%%%%%%%%
for which the algebraic SUSY scheme can be readily applied.
Since $l$ is always an integer for the common spherical harmonics, Eq.~(7)
is moreover a reflectionless one (i.e., the calculated reflection coefficient
is zero).
The energy eigenvalues are known to be of the type $E_{n}=-(l-n)^2$, with
$n=0,1,2...N$, where $N$ is the number of bound states the
$-l(l+1){\rm sech}^2$
potential well can hold, which for spherical harmonics is equal to $l$.
Since $E_{n}=-m^2$, one gets $n=l-m$. The eigenfunctions $|v _{n}(z;0)
\rangle$
can be obtained by repeated application of the creation operators
$A^{\dagger}(z;k)\equiv (-\frac{d}{dz}+k\tanh z)$, $k \in [0, m-1]$, as follows
%%%%%%%%%%%%%%%%%%
$$
|v _{n}(z;0)\rangle\propto \Bigg[\prod _{k=m-1}^{k=0} A^{\dagger}(z;k)
\Bigg]|v_{0}(z;m)\rangle~,
\eqno(3)
$$
%%%%%%%%%%%%%%%%%%
where the ground state wavefunction is
$|v_{0}(z;m)\rangle={\rm sech}^{m} z$.
The connection between the $v$ functions and the associated Legendre
polynomials is given by $P_{l}^{m}({\rm tanh}z)\approx v _{l-m}(z;l)$.
%\approx(1-{\rm tanh}^2 z)^ $.

We notice that the mapping $f$ and the Schr\"odinger equation (7) were already
known to Infeld and Hull \cite{ih}, who used them in their factorization
method.

\bigskip

{\bf III. JACOBI POLYNOMIALS}

%{\bf 3.1}- 
 A more complicated case is that of Jacobi polynomials which
fulfill the differential equation
%%%%%%%%%%%%%%%%%%
$$
(1-x^2)y^{''}+[\beta-\alpha - (\alpha+\beta+2)x]y^{'}+n(n+\alpha+\beta +1)y=0
~.
\eqno(9)
$$
%%%%%%%%%%%%%%%%%%%
By employing the notations
$2(m+1)=\alpha +\beta +2$ , $n(n+\alpha +\beta +1)=(n'-m)(n'+m+1)$,
$\beta - \alpha =\gamma$. Eq.~(9) turns into
%%%%%%%%%%%%%%%%%%%%
$$
(1-x^2)y^{''}+\gamma y^{'}-2(m+1)xy^{'}+(n^{'}-m)(n^{'}+m+1)y=0
\eqno(10)
$$
%%%%%%%%%%%%%%%%%%%%%
and with $v=(1-x^2)^{m/2}y$ one gets
%%%%%%%%%%%%%%%%%%%%%
$$
(1-x^2)v^{''}-(2x-\gamma)v'+
\Big[n^{'}(n^{'}+1)-\frac{m^2-\gamma mx}{1-x^2}\Big]v=0~,
\eqno(11)
$$
%%%%%%%%%%%%%%%%%%%%%%%
or in spherical polar coordinates
%%%%%%%%%%%%%%%%%%%%%%%
$$
\frac{d^2v}{d\theta ^2}+\left({\rm cot}\theta-\frac{\gamma}
{\sin \theta}\right)
\frac{dv}{d\theta}+\Big[n^{'}(n^{'}+1)-\frac{m^2-\gamma m\cos \theta}
{\sin ^2\theta}\Big]v=0~.
\eqno(12)
$$
%%%%%%%%%%%%%%%%%%%%%%%
We already see that Eq.~(12) presents deviations from the standard associated
form.
With the change of variable $\theta=f(z)$ and the notations
$P(f)$ and $Q(f)$ for the coefficients of the first derivative of $v$
and of $v$, respectively, in Eq.~(12), one can get the following form
%%%%%%%%%%%%%%%%
$$
\frac{d^2v}{dz^2}+\Big[f^{'}P(f)-\frac{f^{''}}{f^{'}}\Big]
\frac{dv}{dz}+Q(f)f^{'2}v=0
~.
\eqno(13)
$$
%%%%%%%%%%%%%%%%%%%%
Again the coefficient of the first derivative is put to nought, leading to
the differential equation
%%%%%%%%%%%%%%%%%%%%%%%%%%%
$$
\frac{f^{''}}{f^{'}}=f^{'}P(f)~,
\eqno(14)
$$
%%%%%%%%%%%%%%%%%%%%%%%%%%%
which can be solved by the substitution $u=f^{'}\equiv \frac{df}{dz}$ leading
to
%%%%%%%%%%%%%%%%%%%
$$
\frac{du}{df}=uP(f)~,
\eqno(15)
$$
%%%%%%%%%%%%%%%%%%%%
with the solution $v=e^{\int P(f)df}$. Since
%$\frac{df}{dz}=(1+f)^{\beta +1}(1-f)^{\alpha +1}$.
$\int P(\theta)d\theta=\ln[\frac{\sin \theta}{\tan ^{\gamma}(\theta/2)}]$
we have $\frac{d\theta}{dz}=
\frac{\sin \theta}{\tan ^{\gamma}(\theta/2)}$. Thence,
$z=\int \tan ^{\gamma}(\theta/2){\rm csc}\theta d\theta +{\rm const}$.
We have found that a convenient choice for the constant of integration is
${\rm const}=-\frac{1}{\gamma}$.
Using
$g=\tan (\theta/2)$, one gets $z=\int g^{\gamma -1}dg -\frac{1}{\gamma}$.
Thus,
%%%%%%%%%%%%%%%%%%%%%
$$
z=\left\{\begin{array}{ll}
      \frac{\tan ^{\gamma}(\theta/2)}{\gamma}
                            -\frac{1}{\gamma} & \mbox{$\gamma \neq$ 0}\\
      \ln[\tan(\theta/2)]                     & \mbox{$\gamma$ =0.}
         \end{array}
  \right.
$$
%%%%%%%%%%%%%%%%%%%%%%
The angular variable can be written as
%%%%%%%%%%%%%%%%%%%%%%%
$$
\theta=\left\{\begin{array}{ll}
           2{\rm arctan}[(\gamma z+1)^{1/\gamma}] & \mbox{$\gamma \neq$ 0}\\
             2{\rm arctan}(e^{z})                 & \mbox{$\gamma$ =0.}
         \end{array}
  \right.
$$
%%%%%%%%%%%%%%%%%%%%%%%%%%%%
One can show that the mapping $f$ for $\gamma \neq 0$ is equivalent to the
replacements $\sin \theta =\frac{2}{(\gamma z+1)^{-1/\gamma}+
(\gamma z+1)^{1/\gamma}}={\rm sech} w$ and
$\cos \theta=\frac{(\gamma z+1)^{-1/\gamma}-
(\gamma z+1)^{1/\gamma}}{(\gamma z+1)^{-1/\gamma}+
(\gamma z+1)^{1/\gamma}}=-{\rm tanh}w$, where
$w=\frac{\ln (\gamma z+1)}{\gamma}$.
%The case $\gamma =0$ leads to the Gegenbauer polynomials
%of the previous section.
%On the other hand,
For the general case $\gamma \neq 0$ one can get an equation of the form
%more or less similar to Eq.~(7) as follows
%%%%%%%%%%%%%%%%%%
$$
-\frac{d^2v}{dz^2}-
\Big[n^{'}(n^{'}+1)\frac{{\rm sech}^2(\frac{\ln(\gamma z+1)}
{\gamma})}{(\gamma z+1)^2}
+\frac{m^2+\gamma m\tanh (\frac{\ln(\gamma z+1)}{\gamma})}{(\gamma z+1)^2}
\Big]v
=0~.
\eqno(16)
$$
%%%%%%%%%%%%%%%%%%%
It can be interpreted as a Schr\"odinger equation at zero energy.
One set of solutions $v$ are of the type
$v_{n}=[1-{\rm tanh}^2(\frac{\ln (\gamma z+1)}{\gamma})]
^{\frac{\alpha+\beta}{4}}
P_{n}^{\alpha,\beta}({\rm tanh}(\frac{\ln (\gamma z+1)}{\gamma}))$, where
$P_{n}^{\alpha,\beta}$ are Jacobi polynomials.

%{\bf 3.2}-
 Up to now we did not use the $\alpha,\beta$ asymmetry in the function $v$.
If we do that by using $v=(1-x)^{\alpha/2}
(1+x)^{\beta/2}y$ in Eq.~(10), one gets
%%%%%%%%%%%%%%%%
$$
(1-x^2)v^{''}-2xv^{'}+
\Big[n^{'}(n^{'}+1)-\frac{\delta+\epsilon x}{1-x^2}\Big]v=0~,
\eqno(17)
$$
%%%%%%%%%%%%%%%%%
where $\delta =m(2m+1)+\gamma ^2$ and $\epsilon =-\gamma(3m+1)$. In spherical
coordinates ($x=\cos \theta$) Eq.~(17) reads
%%%%%%%%%%%%%%%%%%
$$
\frac{d^2v}{d\theta ^2}+{\rm cot}\theta\frac{dv}{d\theta}+
\Big[n^{'}(n^{'}+1)-\frac{\delta+\epsilon \cos \theta}{\sin ^2\theta}\Big]v=0~,
\eqno(18)
$$
%%%%%%%%%%%%%%%%%
which though not in the standard associated Legendre form has the advantage
that the change of variable $\theta=2{\rm arctan}(e^{z})$ can be used to
get
%%%%%%%%%%%%%%%%
$$
\frac{d^2v}{dz^2}+[n^{'}(n^{'}+1){\rm sech} ^2z+\epsilon
{\rm tanh}z]v=\delta v~.
\eqno(19)
$$
%%%%%%%%%%%%%%%%
A simplified form containing only the ${\rm tanh}$ function is the following
\cite{dks}
%%%%%%%%%%%%%
$$
\frac{d^2v}{dz^2}+[A{\rm tanh}^2z+2B{\rm tanh}z]v=Cv~,
\eqno(20)
$$
%%%%%%%%%%%%%
where $A=-n^{'}(n^{'}+1)$, $B=-\gamma\frac{(3m+1)}{2}$,
$C=-n^{'}(n^{'}+1)+m(2m+1)+ \gamma ^2$, and it is supposed that $B<n^{'2}$
\cite{dks}.
As discussed by DGS and by other authors \cite{vil} such a Schr\"odinger
equation
corresponds to the motion of an electron in a Coulomb field in the presence
of an Aharonov-Bohm potential and a Dirac monopole potential. It is also
a Rosen-Morse II problem which is known to be shape invariant \cite{dgs}.
The energy eigenvalues and eigenfunctions are in our notations as follows
\cite{dks}
%%%%%%%%%%%%%%%%%%%%%%%%%%
$$
E_n=n^{'}(n^{'}+1)-(n^{'}-n)^2-\frac{B^2}{(n^{'}-n)^2}~,
\eqno(21)
$$
%%%%%%%%%%%%%%%%%%%%%%%%%%
$$
v_{n}=(\sqrt{1-{\rm tanh}z})^{n^{'}-n+r}(\sqrt{1+{\rm tanh}z})^{n^{'}-n-r}
P_{n}^{n^{'}-n+r, n^{'}-n-r}({\rm tanh}z)~,
\eqno(22)
$$
%%%%%%%%%%%%%%%%%%%%
where $r=\frac{B}{n^{'}-n}$.

\bigskip

{\bf IV. GEGENBAUER POLYNOMIALS}

%In this section, we apply the DGS scheme to
The Gegenbauer
polynomials $C_{p}^{q}(x)$ %which
are polynomial solutions of the ultraspherical equation
%%%%%%%%%%%%%%%%%%%%%%%%
$$
(1-x^2)y^{''}-(2q +1)xy^{'}+p(p+2q)y=0~.
\eqno(23)
$$
%%%%%%%%%%%%%%%%%%%%%%%%%%%
%This equation can be put in the following associated Legendre form
%%%%%%%%%%%%%%%%%%%%%%%%%%%%%%%
%$$
%(1-x^2)y^{''}-2(m'+1)xy'+(n'-m')(n'+m'+1)y=0
%\eqno(19)
%$$
%%%%%%%%%%%%%%%%%%%%%%%%%%%%
%by the substitutions $p=n'-m'$ and $q=m'+\frac{1}{2}$. Since in Eq.~(19)
%we want $y$ to be an associated Legendre function, $q$ should be half-integer.
%To Eq.~(19) one can apply the change of function $v=(1-x^2)^{m'/2}y$ to obtain
%the self-adjoint form of the associated Legendre equation
%%%%%%%%%%%%%%%%%%%%%%%%%%%%%%%%
%$$
%(1-x^2)v^{''}-2xv'+\Big[n'(n'+1)-\frac{m'^{2}}{1-x^2}\Big]v=0~,
%\eqno(20)
%$$
%%%%%%%%%%%%%%%%%%%%%%%%%%%%%%%%%
%or in spherical polar coordinates
%%%%%%%%%%%%%%%%%%%%%%%%%%%%
%$$
%\frac{d^2v}{d\theta ^2}+\cot \theta\frac{dv}{d\theta}+
%\Big[n'(n'+1)-\frac{m'^{2}}{\sin ^{2}\theta}\Big]v=0~.
%\eqno(21)
%$$
%%%%%%%%%%%%%%%%%%%%%%%%%
The ultraspherical case can be considered as the particular $\gamma =0$ Jacobi
case in the formulas 9-16 of the previous section, and at the same time as a
slightly generalized spherical harmonics case of the first section.
One gets immediately the shape invariant, exactly solvable
Schr\"odinger equation
%From now on, we can apply the scheme of the previous section, i.e.,
%obtaining the Schr\"odinger form of Eq.~(21) by the change of variable
%$\theta =2\tan ^{-1}(e^{z})$ leading to
%%%%%%%%%%%%%%%%%%%%%%%%%
$$
-\frac{d^2v}{dz^2}-\Big[n'(n'+1){\rm sech} ^2z\Big] v=-m'^{2}v~,
\eqno(24)
$$
%%%%%%%%%%%%%%%%%%%%%%%%%%%%%
where $n^{'}=p+q-\frac{1}{2}$ and $m^{'}=q-\frac{1}{2}$. For a full
equivalence to Eqs.~(6) and (7) $q$ should be half integer.
%Thus, we got a similar shape invariant, exactly solvable Schr\"odinger
%equation, for which one can apply the supersymmetric results \cite{dks}.
The energy eigenvalues of the potential
$V(z)=-n'(n'+1){\rm sech} ^2z$ are $E_{n}=-(n'-n)^2$, $(n=0,1,2,...N)$, where
$N$ is the number of bound states in the potential well, and is equal to the
largest integer contained in $n'$. The eigenfunctions $v _{n}(z;n')$ are
obtained by applying the factorization (creation) operators
$A^{\dagger}(z;a_{i})=(-\frac{d}{dz}+a_{i}\tanh z)$, where $a_{i}=n'-i$,
onto the ground state wave function
$v _0(z;a_{n})={\rm sech} ^{n'-n}z\equiv {\rm sech}^{q-\frac{1}{2}}z$, i.e.,
%%%%%%%%%%%%%%%%%%
$$
v_{n}(z;n')\approx A^{\dagger}(z;n')A^{\dagger}(z;n'-1)
A^{\dagger}(z;n'-2)...A^{\dagger}(z;n'-n+1){\rm sech} ^{n'-n}z~,
\eqno(25)
$$
%%%%%%%%%%%%%%%%%%%%%%%
which is an analog of Eq.~(8).
In the case of this section $n=n'-m'\equiv p$, and therefore we are
dealing with associated Legendre functions of the type
$v_{n}\equiv P_{n'}^{m'}(\tanh z)$. The original
Gegenbauer polynomials can be written formally as
$C_{p}^{q}(x)\propto (1-\tanh ^2z)^{\frac{1-2q}{4}}
P_{p+q-\frac{1}{2}}^{q-\frac{1}{2}}(\tanh z)$; the proportionality is used
since we are not addressing the normalization issue here. Although this
relationship appears to be known, see \cite{wg}, we found it in a simple way.
The $n'(n'+1){\rm sech} ^2z$ potential is reflectionless
if and only if $n'$ is an integer. Since $p$ should be an integer, the
reflectionless property requires again a half integer $q$.

Gegenbauer polynomials are connecting higher-dimensional spherical harmonics
to the usual two-dimensional ones. In the theory of hyperspherical
harmonics they play a role which is analogous to the role played by
Legendre polynomials in the usual three-dimensional space \cite{av}.
Therefore, we believe our discussion to be useful in such a context as well.

\bigskip

\bigskip
{\bf ACKNOWLEDGMENT}

The work was supported in part by
the CONACyT Project No. 4868-E9406 and a CONACyT undergraduate fellowship.
The authors thank the referee for suggestions.

%%%%%%%%%%%%%%%%%%%%%%%%%%%%%%%%%%%%%%%%%%%

\end{document}